\documentclass[sigconf]{acmart} 
\AtBeginDocument{%
	\providecommand\BibTeX{{%
			\normalfont B\kern-0.5em{\scshape i\kern-0.25em b}\kern-0.8em\TeX}}}

\copyrightyear{2022}
\acmYear{2022}
\setcopyright{rightsretained}
\acmConference[SCF '22]{Symposium on Computational
	Fabrication}{October 26--28, 2022}{Seattle, WA, USA}
\acmBooktitle{Symposium on Computational Fabrication (SCF '22),
	October 26--28, 2022, Seattle, WA, USA}
\acmDOI{10.1145/3559400.3562005}
\acmISBN{978-1-4503-9872-5/22/10}

\usepackage{color}
\usepackage{float}
\usepackage{subcaption}
\usepackage{url}
\usepackage{wrapfig}
\usepackage{mathtools}
\usepackage[ruled]{algorithm2e} 

\usepackage{xcolor}

\SetCommentSty{mycommfont}
\usepackage{listings}

	\def\textrightarrow{ -> }%

	\newcommand{\ME}[1]{{#1}}

	\newcommand{\pat}[1]{{\texttt{#1}}}
	
	\def\xR{\mathbb{R}}
	\newcommand{\tin}{\!\in\!}
	
	\def\cm{\mathcal{G}} 
	\def\cmV{\mathcal{S}} 
	\def\cmR{\mathcal{R}} 
	\def\cmC{\mathcal{C}} 
	\def\cmX{X_{\cm}} 
	\def\cmSU{Y_{\cm}} 
	
	\def\cs{s} 
	\def\cw{w} 
	
	\def\nr{N} 
	\def\sad{\Pi} 
	
	\def\mM{M}
	\def\mV{\mathcal{V}}
	\def\mE{\mathcal{E}}
	
	\def\sP{\cs} 
	\def\sW{\cw} 

\newtheorem{thm}{Theorem}[section]
\newtheorem{definition}[thm]{Definition}
	\newtheorem{observation}[thm]{Observation}

\begin{document}

	\title{AmiGo: Computational Design of Amigurumi Crochet Patterns}
	
	
	\author{Michal Edelstein}
	\orcid{0000-0001-9126-1617}
	\affiliation{%
		\institution{Technion - Israel Institute of Technology}
		\city{Haifa}
		\country{Israel}}
	\email{smichale@cs.technion.ac.il}
	
	\author{Hila Peleg}
	\orcid{0000-0002-0107-5659}
	\affiliation{%
		\institution{Technion - Israel Institute of Technology}
		\city{Haifa}
		\country{Israel}}
	\email{hilap@cs.technion.ac.il}
	
	\author{Shachar Itzhaky}
	\orcid{0000-0002-7276-7644}
	\affiliation{%
		\institution{Technion - Israel Institute of Technology}
		\city{Haifa}
		\country{Israel}}
	\email{shachari@technion.ac.il}
	
	\author{Mirela Ben-Chen} 
	\orcid{0000-0002-1732-2327}
	\affiliation{%
		\institution{Technion - Israel Institute of Technology}
		\city{Haifa}
		\country{Israel}}
	\email{mirela@cs.technion.ac.il}
	
	\authorsaddresses{
		Authors' addresses: The Henry and Marilyn Taub Faculty of Computer Science, Technion - Israel Institute of Technology, Haifa, Israel, 3200003}

	\begin{abstract}
		We propose an approach for generating crochet instructions (\emph{patterns}) from an input $3$D model.
		We focus on \emph{Amigurumi}, which are knitted stuffed toys. Given a closed triangle mesh, and a single point specified by the user, we generate crochet instructions, which when knitted and stuffed result in a toy similar to the input geometry.
		Our approach relies on constructing the geometry and connectivity of a \emph{Crochet Graph}, which is then translated into a \ME{c}rochet pattern. 
		We segment the shape automatically into chrochetable components, which are connected using the join-as-you-go method, requiring no additional sewing. We demonstrate that our method is applicable to a large variety of shapes and geometries, and yields easily crochetable patterns.
	\end{abstract}


	
	\begin{teaserfigure}
		\includegraphics[width=\textwidth]{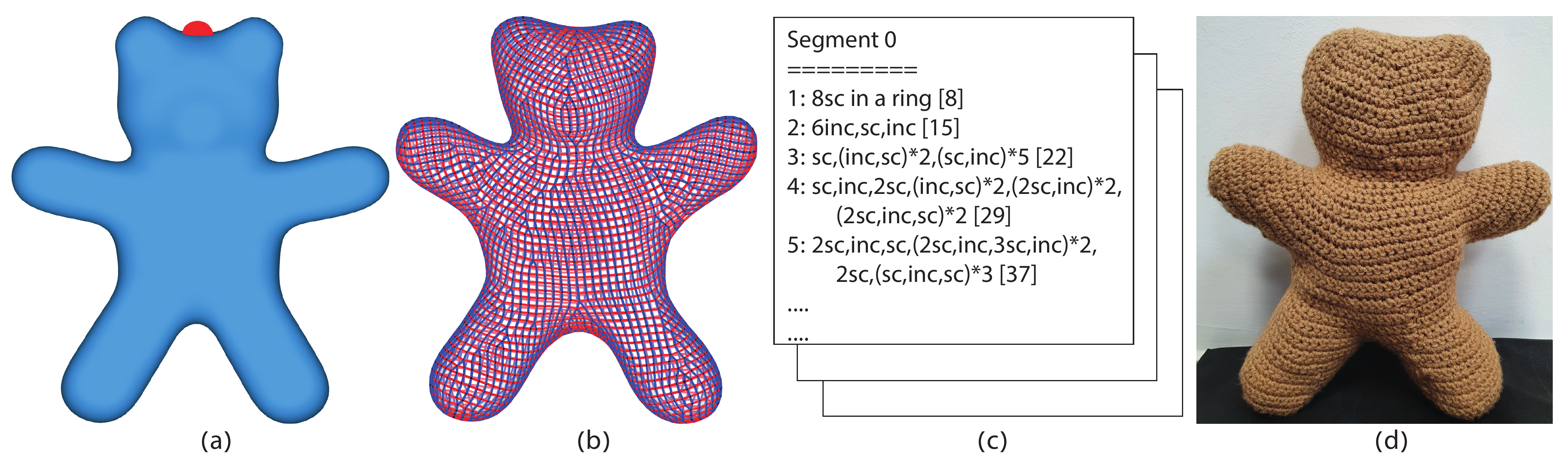}
		\caption{Given an input $3$D model and a seed point (a) we automatically generate a \emph{crochet graph} (b) which is translated into human-readable crochet instructions (c). When crocheted and stuffed, the output is a toy similar to the input shape (d).}
		\label{fig:teaser}
	\end{teaserfigure}
	
	\maketitle
	
	

	\section{Introduction}
	Hand making toys is a popular and ancient craft. Knitting toys and stuffing them is one of the most popular approaches, as it allows for great flexibility in the color and texture of the final product, does not require sophisticated tools, and the technique can be easily taught, even to small children. Knitting approaches are mostly divided into methods which use a single hook or needle, usually classified as \emph{Crochet}, and techniques which use two needles, which are generally called \emph{Knitting}. Despite some superficial similarities, crochet and knitting yield very different fabrics, and the knitting patterns are different as well. 
	
	Computational knitting has been recently investigated in the graphics and fabrication communities~\cite{yuksel2012stitch}, as knitting machines became available and popular. Crochet, on the other hand, cannot be as easily automated, and to-date there is no existing \ME{c}rochet-machine~\cite{seitz2021language}. Furthermore, the research on computational methods for generating \ME{c}rochet patterns is similarly lacking. On the other hand, the crochet community has increased dramatically in recent years~\cite{burns2021happy}, and \emph{Amigurumi}, or crochet stuffed toys, became very popular. As the availability of computational techniques for pattern generation is quite limited, most designers rely on trial and error methods which are tedious and time consuming. Furthermore, novice crocheters are limited to producing existing patterns, and cannot easily express their creativity by generating patterns themselves. 
	
	Our goal in this paper is to address this gap, and suggest a computational method for generating a crochet pattern from an input 3D model. Given a single point on the model chosen by the user, denoted as the \emph{seed}, and a user-selected stitch size, we automatically generate a \emph{crochet graph}. The graph is then used for generating a crochet pattern, which after crocheting and stuffing, resembles the input model. We additionally provide a visualization of the expected crocheted shape, and thus the user can experiment with different seeds and different yarn sizes. We demonstrate that our algorithm is applicable to a large variety of shapes, and compares favorably to prior work. 
	
	
	

	\subsection{Related Work}
	\paragraph{Knitting.} The literature on computational \emph{knitting} (as opposed to \ME{c}rochet) is quite large, and we mention some of it here for completeness.
	Note that, in general, crochet and knitting are two related, but very different, methods of fabric generation, and converting knitting patterns to crochet and vice versa is very challenging, even for human experts. Thus, a computational knitting pipeline cannot be used "as-is" for crochet.
	
	The Stitch Meshes line of work~\cite{yuksel2012stitch} deals with representing 3D models at the yarn level using polygonal meshes whose faces represent different types of stitches, and on top of that considering knitability~\cite{wu2019knittable} and wearability~\cite{wu2021wearable}.
	Our crochet graph representation, on the other hand, is more abstract, representing only stitch heads/bases and stems, instead of the full yarn-level representation. However, for crochet this is sufficient. Specifically, the crochetability constraint of coupled rows, which directly translates into an algorithm for creating the instructions, can be easily enforced on our crochet graph. 
	
	Our representation has some resemblance to the representation of AutoKnit~\shortcite{narayanan2018automatic}, where the nodes represent two stacked stitches and the edges represent the connectivity and stitch size. They build the graph using a user specified \emph{time function} and a set of constraints that guarantee machine knittability. They additionally propose tracing and scheduling algorithms to produce the machine knitting instructions. 
	Popescu et al.~\shortcite{popescu2018automated} build a similar graph by manually dividing the shape into patches and covering them with contours, which are later sampled to produce the instructions.
	Kaspaer at al. \shortcite{kaspar2021knit} also generate machine-knitting instructions based on a stitch graph. The graph is computed by sampling 2D garment patterns according to a specified time function designed by the user.
	Other approaches focus on knitting compilers~\cite{mccann2016compiler}, interactive design of knit templates~\cite{jones2021computational}, and complex knit structures and multi-yarn~\cite{nader2021knitkit}. 
	
	When compared to machine knitting algorithms our approach is better tailored to hand-crocheting. First, the crocheted models are in many cases very low resolution (e.g., if they are aimed for beginners), and therefore we need to adapt our sampling rate accordingly (see Section~\ref{sec:sampling_modification}). Furthermore, short rows are much less common in crochet, and we avoid them to generate patterns which appeal to beginner crocheters. The user only needs to supply an input seed point, instead of constraints on the time function as in AutoKnit, (e.g., two or more seed points), or a manual segmentation as in Popescu et al.~\shortcite{popescu2018automated}. Finally, our simpler approach and the low resolution needed for crochet also leads to shorter processing times, where generating the instructions takes a few minutes, when compared to tens of minutes for AutoKnit. Hence opening the door to interactive editing and design of patterns in future work. 

	\paragraph{Crochet.}
	A recent technical report~\cite{seitz2021language} provides an excellent background about the concept of computational crochet. The authors discuss pattern representation, as well as differences between crochet and knitting, and the lack of crochet machines. 
	One of the first approaches to computational crochet was presented by Igarashi et al.~\shortcite{igarashi2008knitty}, which provided an interactive tool for sketching a model and producing crochet instructions. Later, Nakjan et al.~\shortcite{nakjan2018automatic} suggested a method that allows the user to design dolls using $2$D sketches and generated from them crochet instructions.
	
	Beyond sketching, Igarashi et al.~\shortcite{igarashi2008knitting} also allow the user to start from a $3$D model. There, the mesh is covered with evenly spaced winding strips, which are then sampled at constant distances to compute the pattern. Their method requires a manual segmentation of the input, and the resulting knitted models are often very different visually from the input. More recently, Çapunaman et al.~\shortcite{ccapunaman2017computing} suggested a method that infers the stitch directions and connectivity from the $(u,v)$ parameterization of a given surface. Thus, this approach requires a parameterized surface, whose parameter directions align with the required stitch directions. It is unclear how to achieve such a parameterization for a general surface. Finally, Guo et al.~\shortcite{guo2020representing} extended the Stitch Mesh framework to crochet by defining new faces that represent crochet stitches and new edge types that represent the current loop. They produce crochet instructions for 3D models and simulate the expected geometry. However, the crocheted item can differ greatly from the original model.
	
	There are a few publicly available software projects that allow users to generate instructions from very simple geometries. The Crochet Sphere Calculator~\cite{sphereCalc} generates instructions for crocheting spheres with a given number of rows. The Crochet \ME{L}athe \cite{crochetLathe} generates crochet instructions for surfaces of revolution, by allowing the user to design the profile curve. These are very basic, and do not allow the user to input a general $3$D model.

	\begin{figure*}[t]
		\centering
		\includegraphics[width=.8\linewidth]{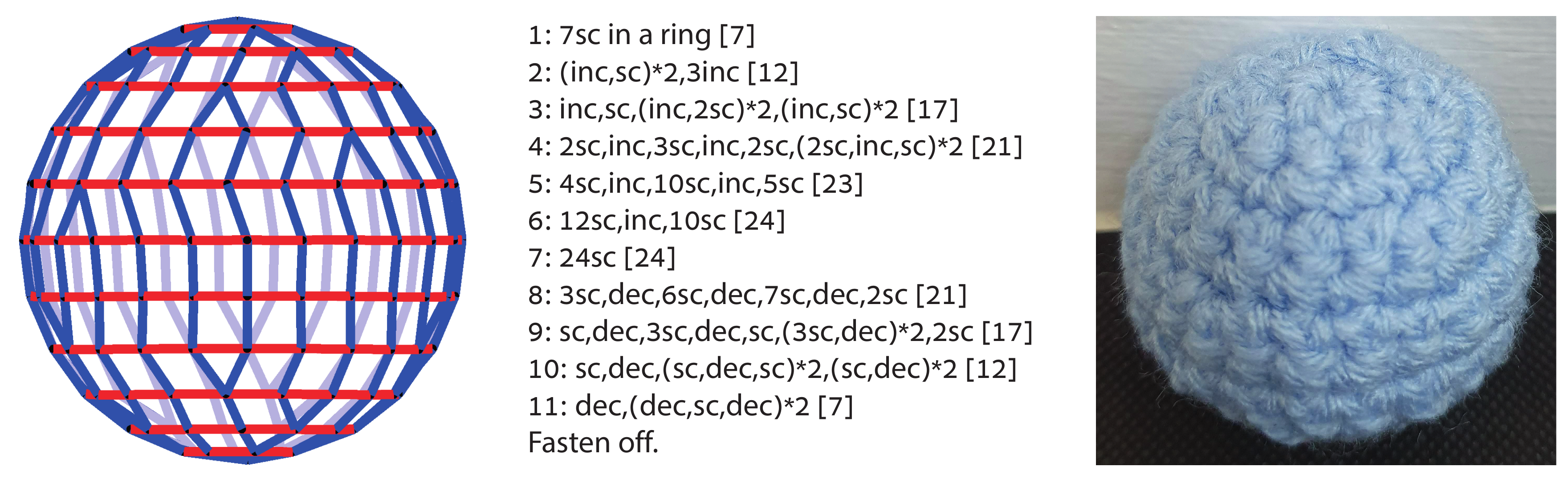}
		\caption{(left) The \emph{Crochet Graph} of a sphere, with row edges $\cmR$ in red, and column edges $\cmC$ in blue. (middle) The corresponding instructions (pattern) for crocheting the sphere. (right) The crocheted sphere.} 
		\label{fig:crochet_graph}
	\end{figure*}

	\subsection{Contribution}
	We propose an automatic method for generating crochet instructions (patterns) from a closed input $3$D model, a seed point, and a stitch size. The generated instructions are crochetable, use only simple crochet stitches, and are based on the "join-as-you-go" method, and thus do not require any sewing. When crocheted and stuffed the models are similar to the input $3$D shape, much more so than any previous approach for crochet instructions generation.

	\section{Representation}
	
	\subsection{Background and Notations}
	
	Given a closed manifold triangle mesh $\mM=(\mV,\mE)$, a \emph{seed} vertex $\cs\tin V$, and a stitch width $\cw\tin\xR$, our goal is to generate human-readable instructions $P(\mM,\cs,\cw)$ for crocheting $\mM$ from the point $\cs$, with the given stitch width $\cw$. 
	
\begin{figure}[t]
		\centering
		\includegraphics[height=0.2325\linewidth]{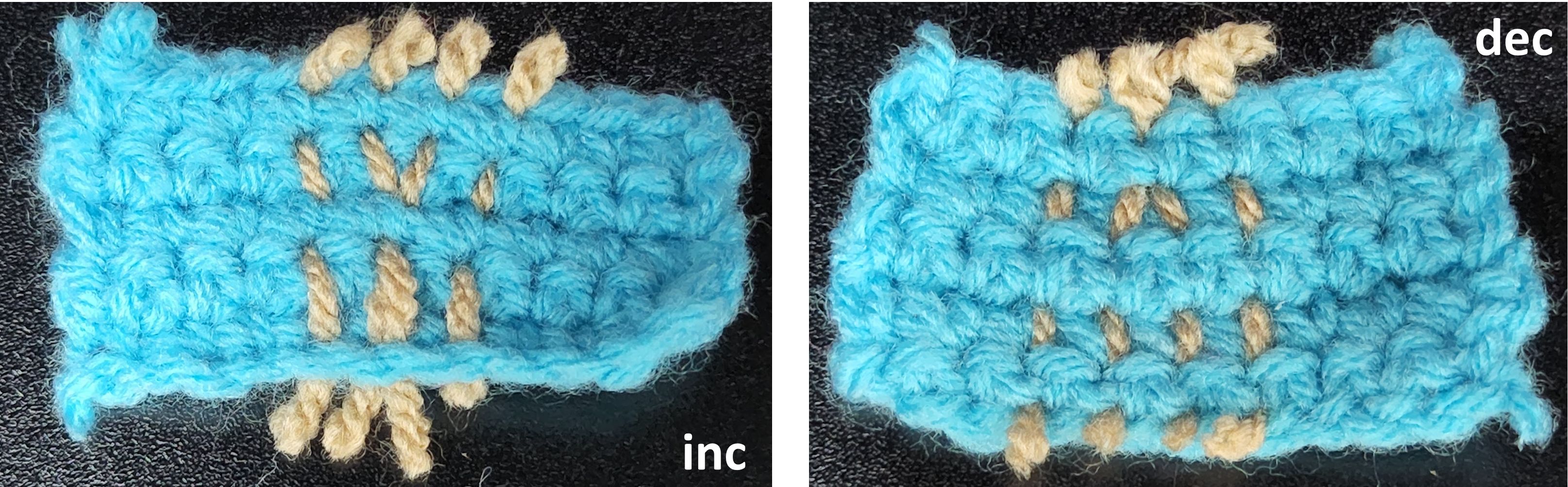}
		\hspace{.0002\linewidth}
		\includegraphics[height=0.2325\linewidth]{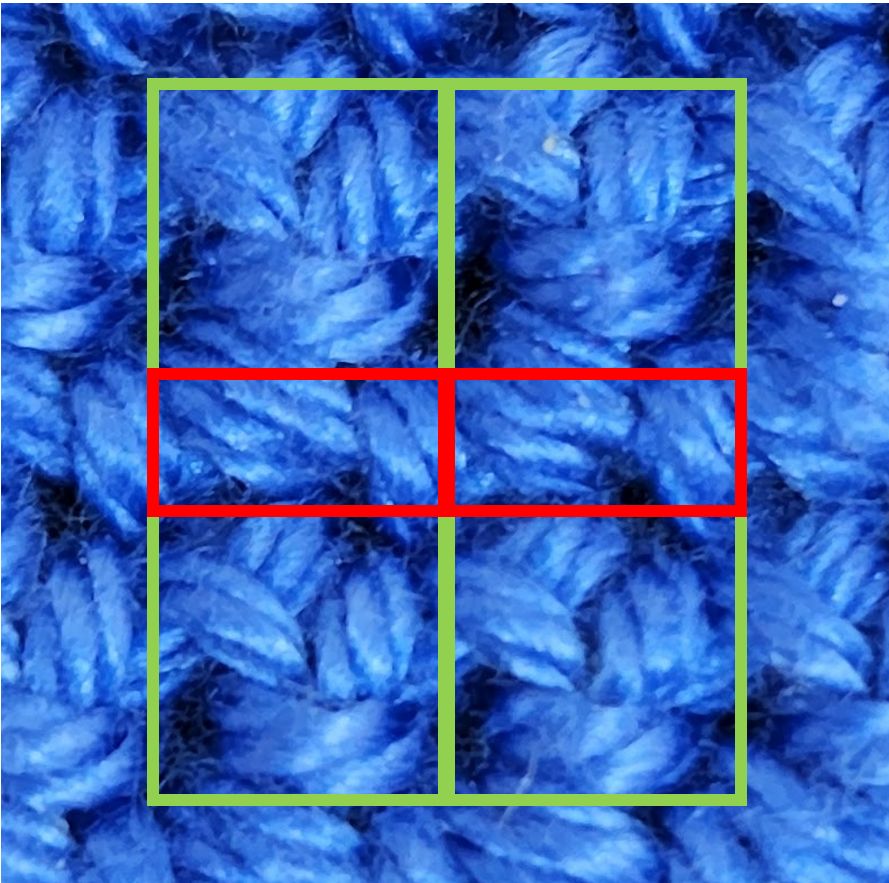}
		\caption{(left) Crochet stitches \pat{inc,dec} marked using yarn on a crocheted patch. (right) The anatomy of crochet stitches, marked are the \emph{top}/\emph{bottom} of the stitch in red and the \emph{stem} in green.} 
		\label{fig:stitches}
	\end{figure}

	Crochet has a wide variety of stitches, and we focus here on the  simple stitch used for Amigurumi, named \emph{single crochet} (\pat{sc}). This is an approximately square stitch, thus covering $M$ with \pat{sc} stitches is equivalent to constructing a quad re-mesh of $\mM$, where each quad is a square, and all the edge lengths are constant.
	This is of course not possible unless the surface is developable, i.e. has zero Gaussian curvature. In practice, curved geometry is accommodated in crochet by introducing stitches which locally \emph{increase} (\pat{inc(x)}) or \emph{decrease} (\pat{dec(x)}) the amount of stitches by $x$. Figure~\ref{fig:stitches} (left) shows an example of the \pat{inc} and \pat{dec} stitches on a crocheted patch. 
	Crochet instructions for Amigurumi typically include \emph{rows}, where each row is a series of \pat{sc, inc, dec} stitches. Figure~\ref{fig:crochet_graph} (\ME{middle}) shows the instructions (pattern) for crocheting the sphere in Figure~\ref{fig:crochet_graph} (\ME{right}). 
	
	\subsection{The Crochet Graph}
	A crochet stitch is composed of a \emph{top}, a \emph{base} and a \emph{stem}, where the \pat{inc,dec} stitches have multiple stems, see Figure~\ref{fig:stitches} (right). 
	The top of one stitch is always the base of some stitch on the next row, and similarly, each stitch has a base on the previous row.
	Therefore, a natural abstraction of the stitch pattern is to consider the stitches and their interconnections as a \emph{graph}. 
	
	Specifically, we define the \emph{Crochet Graph} $\cm=(\cmV,\cmR \cup \cmC)$, whose vertices $\cmV$ are tops/bases of stitches, where a vertex $(i,j) \tin \cmV$ is the base of the $j$-th stitch in row $i$, and the vertices in each row are consecutively ordered. The \emph{column} edges $\cmC$ are stems of stitches, and the connectivity between the bases in each row is represented by the \emph{row} edges $\cmR$. We denote the total number of rows by $\nr$.
	Figure~\ref{fig:crochet_graph} (left) shows the crochet graph corresponding to the crocheted sphere in Figure~\ref{fig:crochet_graph} (right).

	A crochet graph is an intermediate representation between the input triangle mesh $\mM$, and the output instructions $P$. Our goal is to generate a graph such that it is (1) translatable to valid crochet instructions $P$, and (2) when $P$ is crocheted and stuffed, the result resembles the input mesh. Note that there exist multiple instructions $P$ for the same graph $\cm$, and within this space we aim for instructions which are \emph{human-readable}. 
	
	We base our algorithm on the following observations.
	
	\begin{definition}
		A \emph{coupling}~\cite{gold2018dynamic} $C = (c_1,..,c_k)$ between two sequences $A=(p_1,..,p_n)$ and $B=(q_1,..,q_m)$ is an ordered sequence of distinct pairs of points from $A \times B$, such that $c_1 = (p_1, q_1), c_k = (p_n, q_m)$ and 
		\begin{equation}
			c_r = (p_s, q_t) \Rightarrow c_{r+1} \in \big\{(p_{s+1},q_t), (p_s, q_{t+1}), (p_{s+1},q_{t+1})\big\}, \quad \forall r < k.
		\end{equation}
	\end{definition}
	
	\begin{definition}
		Let $\cmV_i,\cmV_{i\!+\!1}, 1 \leq i < \nr$, be the vertices of two consecutive rows of $\cm=(\cmV,\cmR \cup \cmC)$, where $\cmV_i = \big((i,1),..,(i,n_i)\big)$, where $(i,j)\tin\cmV$, and $n_i$ is the number of vertices in row $i$. If there exists a coupling $C$ between $\cmV_i$ and $\cmV_{i\!+\!1}$ such that for all $p_s\tin\cmV_i, q_t\tin\cmV_{i+1}$ we have that $(p_s,q_t)\tin C$ if and only if $(p_s,q_t)\tin\cmC$, then the two rows are \emph{coupled}.
	\end{definition}
	
	\begin{observation} 
		If all the pairs of consecutive rows of $\cm$ are coupled, then there exist valid crochet instructions $P(\cm)$ that use only the instructions \pat{sc}, \pat{inc(x)} and \pat{dec(x)}.
		\label{thm:grid_graph}
	\end{observation}
	
	
	\begin{definition}
		Let $\cmX:\cmV\to\mM$ be an embedding of the vertices of $\cm$ on $\mM$. An \emph{embedded edge} of $\cm$ is a shortest geodesic between the embedding of two vertices of $\cm$ which share an edge, or between the embedding of the first and last vertices on the same row.
	\end{definition}
	
	\begin{definition}
		Let $(p,q)\tin\mM$ be two points whose geodesic distance is larger than some constant that depends on the stitch width $\cw$, and let $\gamma_{p,q}$ be the shortest geodesic between them. If $\gamma_{p,q}$ intersects some embedded edge of an embedding $\cmX$, for any two such points, then we say that $\cmX$ \emph{covers} $\mM$.
	\end{definition}
	
	\begin{observation}
		Let $\cmX:\cmV\to\mM$ be an embedding of the vertices of $\cm$ which covers $\mM$, and $P(\cm)$ valid crochet instructions for $\cm$.
		If all the edge lengths induced by $\cmX$ are equal to $\cw$, then when $P(\cm)$ will be crocheted and stuffed the result will be "similar" to $\mM$.
		\label{conj:cm_embedding}
	\end{observation}

	We discuss Observation~\ref{thm:grid_graph} in section~\ref{sec:graph_to_program}, where we show how to translate the graph into valid instructions. The Observation~\ref{conj:cm_embedding} is in fact true only for a subset of meshes, as we discuss in the next section.
	
	
	
	\subsection{Crochetable Models}
	\paragraph{Curvature.} Crocheting the patterns yields an empty flexible shell of fabric, which obtains its final shape by \emph{stuffing}. 
	Whether the stuffed model obtains the intended shape depends on how the model is stuffed (lightly, firmly, uniformly), as the yarn has some flexibility and will extend to accommodate if the model is over-stuffed. We will assume that the model is stuffed enough to attain maximal volume, but not too much to cause the yarn to stretch and generate gaps. Thus, we expect the resulting stitch size to be similar to the edge lengths induced by the embedding of the crochet graph. If for a given graph $\cm$ its embedding in $3$D with edge lengths $\cw$ is \emph{unique}, we expect the crocheted and stuffed shape to be similar to the input surface.
	
	Importantly, unless the shape is convex, the edge lengths alone (i.e., the \emph{metric}) do not contain enough information to uniquely determine the shape of a non-stuffed model. For example, a surface that has a "crater" leads to edge lengths which can be realized either as a crater or as a "hill". However, if we add the maximal volume assumption, only the hill is possible. This in fact implies that surfaces which have "craters", or more formally, regions of negative mean curvature with \emph{positive} Gaussian curvature, cannot be realized by crocheting and stuffing alone. This is similar to the observation made by Konakovich et al.~\shortcite{konakovic2018rapid}, that surfaces which have negative mean curvature cannot be realized by maximizing the volume with a given conformal metric (i.e., only isotropic scaling is allowed relative to the flat configuration). We handle this case similarly, by preprocessing the model so that it does not contain "craters" (Section~\ref{sec:preprocessing}).
	
	Furthermore, in our case, since we allow anisotropic scaling, negative mean curvature with \emph{negative} Gaussian curvature is in fact possible, but requires a modified sampling rate, which we discuss in Section~\ref{sec:sampling_modification}. To conclude, in terms of geometric obstructions to crochetability, the four possible curvature situations are summarized in Table~\ref{tab:crochetability_curvature}. Note that, as with any sampling-dependent method, if the sampling rate (namely, the number of rows $\nr$) is too small compared to the feature size of the model, the crocheted output will lose some of the geometric detail.
	
	\begin{table}[]
		\centering
		\caption{Curvature obstructions to crochetability, see the text for details.}
		\vspace{-10pt}
		\begin{tabular}{c|c|c}
			\raisebox{0ex}{\hspace{15.5ex}Mean} &   & \\
			\raisebox{0ex}{\hspace{11ex}Curvature} &   &  \\
			\raisebox{0ex}{\hspace{-11ex}Gaussian}  & Positive & Negative \\
			\raisebox{0ex}{\hspace{-10ex}Curvature} &  &  \\
			\hline
			Positive & Crochetable & Preprocessing \\
			Negative & Crochetable & Sampling modification
		\end{tabular}
		\label{tab:crochetability_curvature}
	\end{table}
	
	\paragraph{Branching.} 
	Observation~\ref{conj:cm_embedding} requires that the crochet graph embedding $\cmX$ covers the input surface $\mM$. Because of the special structure of this graph, this induces additional constraints on the possible geometries. Intuitively, models that branch (see Figure~\ref{fig:homer}) cannot be covered in this way. Mathematically, this means that the geodesic distance function on $\mM$ from the embedding of the seed vertex $\cs$ cannot have saddles. This is solved by segmenting the shape, and crocheting the segments in an iterative manner. We explain this in detail in Section~\ref{sec:multiple_segments}.

	\vspace{\baselineskip}
	We first explain the generation of the crochet pattern for a simple non-branching model with positive mean curvature, and then discuss how we handle negative mean curvature and branching.
	
	\section{Overview}
	
	Given a 3D mesh $\mM$, a seed point $\cs$ and a stitch width $\cw$, we first compute a crochet graph $\cm$ and its embedding $\cmX$ such that they adhere to Observations~\ref{thm:grid_graph} and~\ref{conj:cm_embedding} (Section~\ref{sec:mesh2graph}). Then we compute the crochet pattern $P(\cm)$ (Section~\ref{sec:graph2pattern}).
	
	To generate $\cm$ and $\cmX$, we first compute the embedding of the vertices $\cmV$ on $\mM$ (Section~\ref{sec:geometry}), and then derive from that the connectivity of $\cm$, i.e. the row edges $\cmR$ and column edges $\cmC$ (Section~\ref{sec:connectivity}). To compute the pattern $P(\cm)$, we first translate the graph into a program using standard code synthesis tools (Section~\ref{sec:graph_to_program}), and then apply loop unrolling to make the pattern human-readable (Section~\ref{sec:program_to_pattern}). See Algorithm~\ref{alg:main_alg}.
	
	

	

	\begin{algorithm}[b]
		\SetAlgoLined
		\KwIn{A triangle mesh $\mM$, seed $\sP$, stitch width $\sW$}
		\KwOut{Embedded crochet graph $\cm=(\cmV,\cmR\cup\cmC),\cmX$, \ME{c}rochet pattern $P(\cm)$}
		\textbf{Mesh to Graph} \tcp*[l]{Section~\ref{sec:mesh2graph}} \
		\Indp
		Geometry $\cmV, \cmX$ \tcp*[l]{Section~\ref{sec:geometry}} \
		Connectivity $\cmR, \cmC$  \tcp*[l]{Section~\ref{sec:connectivity}} \
		\Indm
		\textbf{Graph to pattern} \tcp*[l]{Section~\ref{sec:graph2pattern}} \
		\Indp
		Graph to program \tcp*[l]{Section~\ref{sec:graph_to_program}} \
		Program to pattern $P(\cm)$ \tcp*[l]{Section~\ref{sec:program_to_pattern}} \
		\Indm
		\caption{An outline of our algorithm}
		\label{alg:main_alg}
	\end{algorithm}
	
	\section{Mesh to Crochet Graph}
	\label{sec:mesh2graph}
	
	\begin{figure*}[t]
		\centering
		\includegraphics[width=1\linewidth]{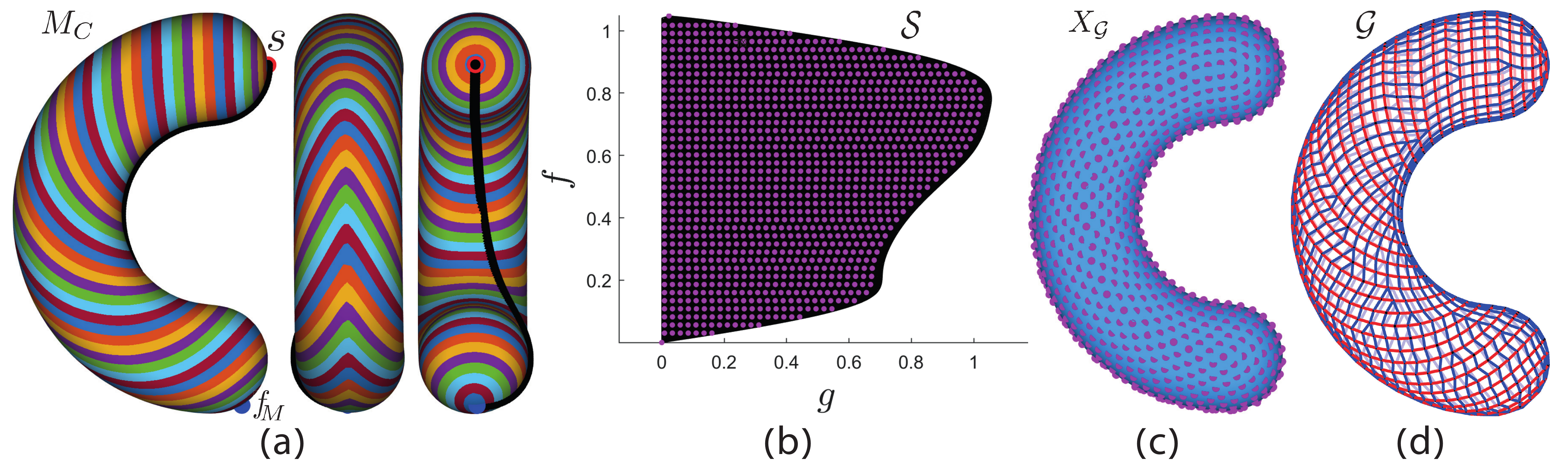}
		\caption{(a) The isolines of $f$, with the seed (red), the maxima of $f$ (blue), and the cut (black) marked. (b) The $f,g$ parameterization, and the sampled grid $\cmV$, (c) The pushed forward points $\cmX$. (d) The output crochet graph $\cm$, with red row edges $\cmR$ and blue column edges $\cmC$.} 
		\label{fig:mesh2graph_c}
	\end{figure*}
	
	\subsection{Geometry}
	\label{sec:geometry}
	Observation~\ref{thm:grid_graph} implies that the vertices $\cmV$ should be grouped into ordered rows, where in each row the vertices have a well defined order. We address this requirement by computing two non-negative monotonically increasing, constant speed functions $f,g:\mM\to\xR$ which define the row-order and column-order of every point on $\mM$.
	Furthermore, Observation~\ref{conj:cm_embedding} implies that the distance between embedded rows, and between embedded vertices in the same row should be $\cw$. We address this by sampling $f,g$ appropriately.
	
	\paragraph{Row order $f$}
	Our models are closed (so they can be stuffed), and therefore the first and last rows in the graph $\cm$ contain a single vertex. The first row contains only the seed $\cs$, and its function value is $f(\cs)=0$. We take $f(v), v\tin\mV$ to be $f(v)=d(v,\cs)$, where $d$ is the geodesic distance. Thus, the isolines of $f$ are rows, and two points $p,q\tin\mM$ are on the same row if $f(p)=f(q)$. 
	If $f$ has more than one maximum, then we need to handle branching (see Section~\ref{sec:multiple_segments}). Otherwise, the vertex that attains the maximum of $f$, denoted as $f_M$ will be the single vertex on the last row.
	
	\paragraph{Column order $g$}
	We first cut $\mM$ along a geodesic from $\cs$ to $f_M$, so that our model and the graph that we compute have the same topology, and denote the cut model by $\mM_C$. The requirements are that \emph{within each row} the vertices of $\cm$ have a well defined order. A row is an isoline of $f$, and therefore the rate of change along the isoline is given by the directional derivative of $g$ in the direction of the tangent to the isoline. Specifically, the tangent to the isoline of $f$ at a point $p\tin\mM$ is given by $J \nabla f$, where $J$ is the rotation by $\pi/2$ in the tangent plane of $p$. Thus to find $g$, we solve an optimization problem whose objective is to minimize $\int_{\mM_C} |\langle J \nabla f, \nabla g \rangle - 1|^2$, s.t. $g(\mathcal{B}) = 0$. Here, $\mathcal{B} \subset \mV$ is the longest connected path of boundary vertices of $\mM_C$ along which $f$ is strictly monotone. 


	
	
	

		

	\paragraph{Sampling}
	The functions $f,g$ define a parameterization of $\mM_C$ to the plane. We conjecture that this parameterization is bijective (as it was in all of our experiments), but leave the proof to future work. The parameterization may have a large metric distortion, however, if $f(p)=f(q)=f_0$ for some two points $p,q\tin\mM$, then $|g(p)-g(q)|$ is equal to the length of the isoline of $f_0$ between $p$ and $q$. 
	Therefore, we uniformly sample $f,g$ on a $2D$ grid of width $\cw$, yielding the vertices of $\cmV$ with indices $(f/\cw, g/\cw)$. Pushing forward the sampled points to the mesh $\mM_C$ yields the embedding of $\cmV$ on $\mM_C$ (and therefore $\mM$), namely $\cmX$.


	\subsection{Connectivity}
	\label{sec:connectivity}
	\paragraph{Row edges $\cmR$}
	Each two consecutive vertices of $\cmV$ on the same row are connected by a row edge. Namely, $\cmR = \bigcup_{i=1}^\nr \cmR_i$, and $\cmR_i = \big\{\big((i,j),(i,j\!+\!1)\big)\,|\,j\tin\{1,..,n_i\!-\!1\}\big\}$. Here $n_i = \big|\cmV_i\big| = \big|\{(i,j)\tin\cmV\}\big|$, namely the number of vertices in the $i$-th row. 
	
	Let $x,y\tin\cmV$ be two consecutive vertices on the $i$-th row. Then we have that $f(x)=f(y)=f_0$ and $|g(x)-g(y)| = \cw$. Therefore, $d_{\gamma({f_0})}\big(\cmX(x),\cmX(y)\big) = \cw$, where $\gamma(f_0)$ is the isoline of $f_0$ on $\mM$, and $d_{\gamma({f_0})}$ is the distance along the isoline. 
	Hence, the Euclidean distance between the embedded vertices $||\cmX(x) - \cmX(y)|| \leq \cw$, and the distance tends to $\cw$ for a "small enough" stitch size. Here, "small enough", means on the order of the square root of the radius of curvature of $\gamma(f_0)$, which is given in terms of the normal curvature in direction $J \nabla f$.
	
	\paragraph{Column edges $\cmC$}
	First, Observation~\ref{thm:grid_graph} requires that all pairs of consecutive rows are coupled. Let $C_i$ be the coupling corresponding to rows $\cmV_i, \cmV_{i+1}$, and let $(p_s,q_t)\tin C_i$. Since $p_s$ and $q_t$ are on consecutive rows, and therefore embedded on isolines of $f$ which differ by $\cw$, the minimal distance $||\cmX(p_s)-\cmX(q_t)||$ is close to $w$. 
	Therefore, if among all couplings we seek the minimizer of:
	\begin{equation}
		\min_{C_i:coupling} \sum_{(p_s,q_t)\in C_i} ||\cmX(p_s)-\cmX(q_t)||,
	\end{equation}
	then the length of the column edges will be close to $\cw$.
	
	A minimal coupling between every pair of consecutive rows is found by Dynamic Time Warping (DTW)~\cite{sakoe1978dynamic,gold2018dynamic}.
	
	
	
	\section{Crochet Graph to instructions}
	\label{sec:graph2pattern}
	
	\begin{figure*}[t]
		\centering
		\includegraphics[draft=false,width=.8\linewidth]{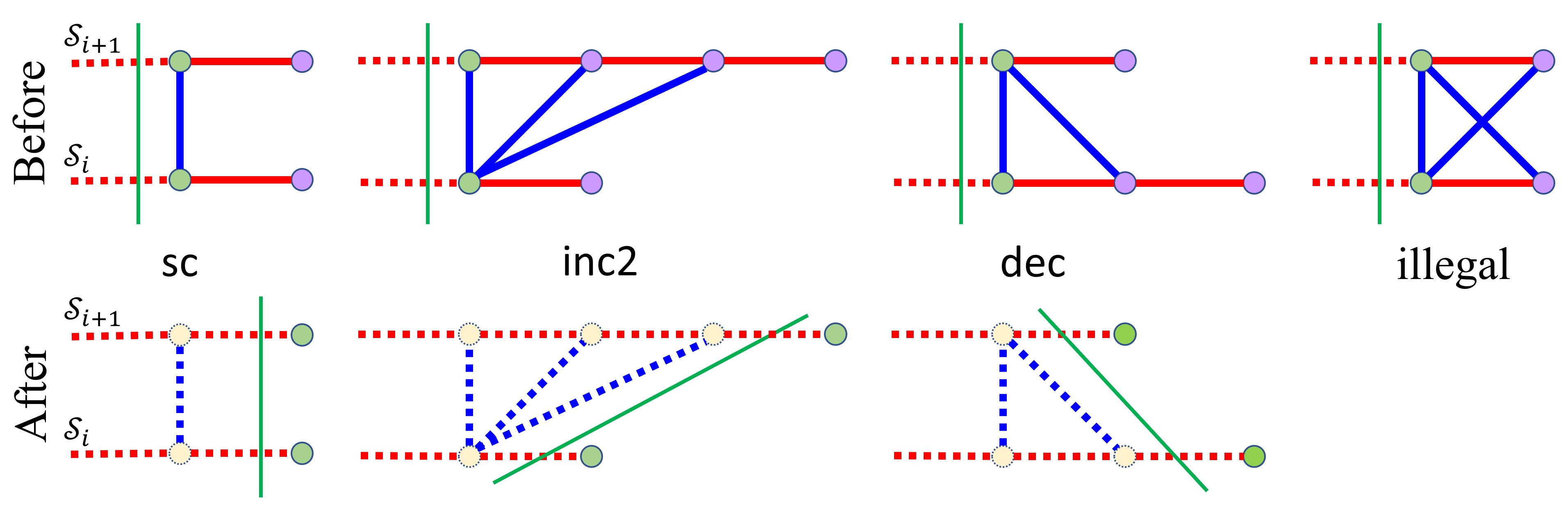}
		\caption{The state of the transducer before (top) and after (bottom) producing a stitch. Vertices at the head of the rows are marked in green, unconsumed vertices in purple, and consumed vertices in yellow.
			See the text for details.
		}
		\label{fig:transducer}
	\end{figure*}
	
	\subsection{Graph to Program}
	\label{sec:graph_to_program}
	In order to turn the \ME{c}rochet graph into instructions, we rely on the following observation: crochet instructions (patterns) constitute an \emph{instruction set}, and as such, crocheting is an \emph{execution} and the finished object is an \emph{execution result}. Moreover, because of the nature of a crocheted object, it is not only a result but a step-by-step \emph{execution trace}.
	
	Therefore, given a crochet object, or in this case its graph representation $\cm$, deriving instructions constitutes a form of \emph{execution reconstruction}~\cite{zuo2021execution},
	a method for reconstructing the set of instructions that lead to an execution trace.
	While reconstructing an execution trace generally requires searching an exponential space of possible instruction sequences,
	the crochet instruction set is limited enough that reconstituting a trace is done in linear time using a transducer.
	
	
	In order to reconstruct the trace, the degree of the graph vertices in both the current row $\cmV_i$ and next row $\cmV_{i+1}$ must be considered.
	Therefore, an execution reconstruction transducer accepts two consecutive rows and their connecting edges, and advances on the rows based on the vertex degrees in either row.
	For example, if at $\cmV_{i+1}$, the transducer is at a vertex with one edge connecting it to the previous row $\cmV_i$, and at the current head of $\cmV_i$ there is only a connection to the current head of $\cmV_{i+1}$, the transducer will yield a \pat{sc}.
	If there are $x>1$ connected vertices in $\cmV_i$ to the head of $\cmV_{i+1}$, the transducer will advance on $\cmV_i$ to consume them all, advance on a single vertex on $\cmV_{i+1}$, and produce a \pat{dec(x)}. Analogously, for $x>1$ connected vertices in $\cmV_{i+1}$ to the head of $\cmV_i$, the transducer will advance on $\cmV_{i+1}$ to consume them, advance on a single vertex on $\cmV_i$ and then produce a \pat{inc(x)}. Figure~\ref{fig:transducer} (left, middle) shows an illustration of the different situations.
	This is entirely analogous to the way crocheting the row is done.
	
	Because the transducer chooses its transition based on which vertex at the current heads of the rows has a degree larger than $1$, it is not technically deterministic.
	However, for a non-deterministic choice to exist, both the head of $\cmV_i$ should be connected to multiple vertices in $\cmV_{i+1}$ and vice versa. This is impossible, since all the pairs of consecutive rows of $\cm$ are \emph{coupled} (see Figure~\ref{fig:transducer}(right)).
	Thus, \emph{under the constraints of the input}, i.e. a valid crochet graph $\cm$,
	all the transitions are mutually exclusive, rendering the transducer's behavior essentially deterministic.

	\subsection{Program to Human-readable Pattern}
	\label{sec:program_to_pattern}
	The instructions in a reconstituted trace can get quite repetitive. The raw output can contain the following rows:
	\begin{verbatim}
		row 2: sc, inc, sc, sc, sc, inc, sc, sc, sc, inc, sc, sc
		row 3: sc, inc, sc, sc, sc, inc, sc, sc, sc, inc, sc, sc
	\end{verbatim}
	indicating that \pat{row 2} is constructed via an \pat{sc} instruction followed by an inc then another two \pat{sc}, repeating three times, then \pat{row 3} is constructed the same way. 
	To make the instructions both succinct and more human-readable, it is customary to convert this code to:
	\begin{verbatim}
		rows 2-3: (sc, inc, 2sc)*3 
	\end{verbatim}
	In order to do this, we employ a disassembly technique called \emph{loop folding}~\cite{lee1994transformation},
	which finds maximal repetitions of instructions and turns them into loops.
	Loop folding of crochet instructions occurs at three different levels: repeating rows, repeating sequences of stitches, and repeating stitches.
	In order to find maximal repeating sequences, the order in which loops are folded must be: sequences first, and then repeating stitches. In the above example, if repeating stitches were folded first, \pat{row 2} after this initial folding would be \pat{sc, inc, 3sc, inc, 3sc, inc, 2sc}, which means the repetition that can be identified within the row would have been smaller. Instead it is first folded to find the repeating sequence \pat{(sc, inc, sc, sc)*3}, which is maximal, and then the internal sequence is re-folded to identify the \pat{2sc}. Finally, identical rows are folded together.
	
	\section{Crochet graph to $3$D embedding}
	\label{sec:graph2embedding}
\begin{figure*}[t]
		\centering
		\includegraphics[width=\linewidth]{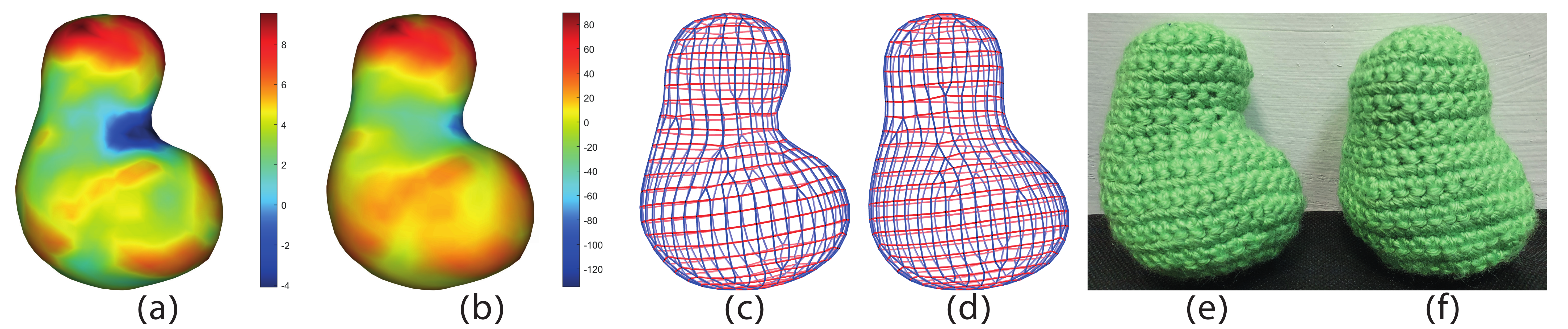}
		\caption{A model with negative mean curvature (a) and negative Gaussian curvature (b) in the same region. The corresponding embeddings $\cmSU$ (c,d) and knitted objects (e,f), with curvature-adapted (c,e) and uniform  (d,f) sampling rate of $g$. Note that the uniform sampling rate does not lead to a model that is similar to the input, whereas curvature-adapted sampling yields a better result.}
		\label{fig:leg_a}
	\end{figure*}

	We generate a $3$D embedding $\cmSU$ of the crochet graph $\cm$ using ShapeUp~\cite{bouaziz2012shape}, in order to obtain a visualization of the expected result. 
	Interestingly, using purely geometric conditions the expected result is quite similar to the crocheted model in practice.
	We use the following constraints for ShapeUp: (1) The column edges which represent \pat{sc} stitches, as well as the row edges are constrained to have length $\cw$, (2) the embedding of the seed point $\cs$ is fixed to the position of the seed vertex, and (3) smoothness.
	We initialize $\cmSU$ using the sampled points $\cmX$. Some of the figures (see for example Figure~\ref{fig:leg_a}) show in addition to the embedded crochet graph $\cmX$ the ShapeUp result $\cmSU$. 
	
	\section{Obstructions to crochetability}
	
	\begin{figure*}[t]
		\centering
		\includegraphics[width=\linewidth]{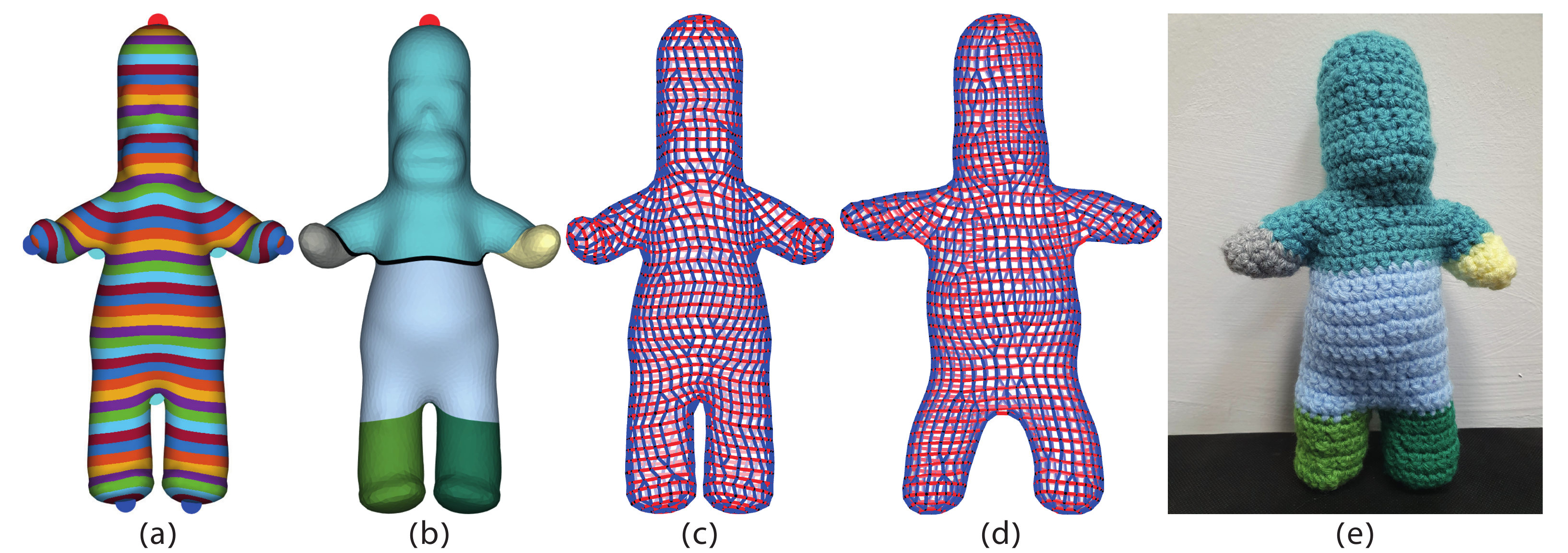}
		\caption{(a) The isolines of $f$ for the marked red seed point, and the saddles (cyan) and maxima (blue) of $f$. (b) The resulting segments. (c) The crochet graph $\cm,\cmX$. (d) The embedding $\cmSU$ of the crochet graph. (e) The final knitted model. Different segments were crocheted in different colors for better visualization.}
		\label{fig:homer}
	\end{figure*}
	
	\subsection{Negative mean curvature}
	As curvature computations are not scale invariant, all the models are normalized to have surface area equal to $1$, so that we can use the same parameters across all the models.
	
	\subsubsection{Positive Gaussian curvature}
	\label{sec:preprocessing}
	Crocheting and stuffing a model which has "craters", i.e. regions of positive Gaussian curvature and negative mean curvature, will not yield a geometry that is similar to the input mesh. Thus, we apply a pre-processing step (similarly to Konakovic et al.~\shortcite{konakovic2018rapid}) to smooth out the craters. Specifically, we apply Conformal Mean Curvature Flow~\cite{kazhdan2012can} localized to these areas until the mean curvature is positive everywhere.
	
	\subsubsection{Negative Gaussian curvature}
	\label{sec:sampling_modification}
	The sampling rate of the isolines of $f$ is determined by the directional derivative of $g$ w.r.t. the tangent to the isoline, namely by $\langle \nabla g, J \nabla f \rangle$. If the curvature in the direction of the isoline $k_{J \nabla f}$ is large compared to the stitch width $\cw$, a uniform sampling rate is inadequate, and does not result in a similar geometry. We therefore adjust the sampling rate in these regions by setting $\langle \nabla g, J \nabla f \rangle = h(k_{J \nabla f})$. We take $h(x) = \tanh(-x/\alpha)/2+1$, to avoid degenerate sampling rates, with $\alpha=10$. Figure~\ref{fig:leg_a} shows an example of such a model. In the central region, the model has negative mean and Gaussian curvature (a,b). Using a uniform sampling rate does not fully reconstruct the negative curvature (d,f), whereas a curvature adapted sampling does (c,e).

	

	\begin{figure*}[t]
		\centering
		\includegraphics[width=\linewidth]{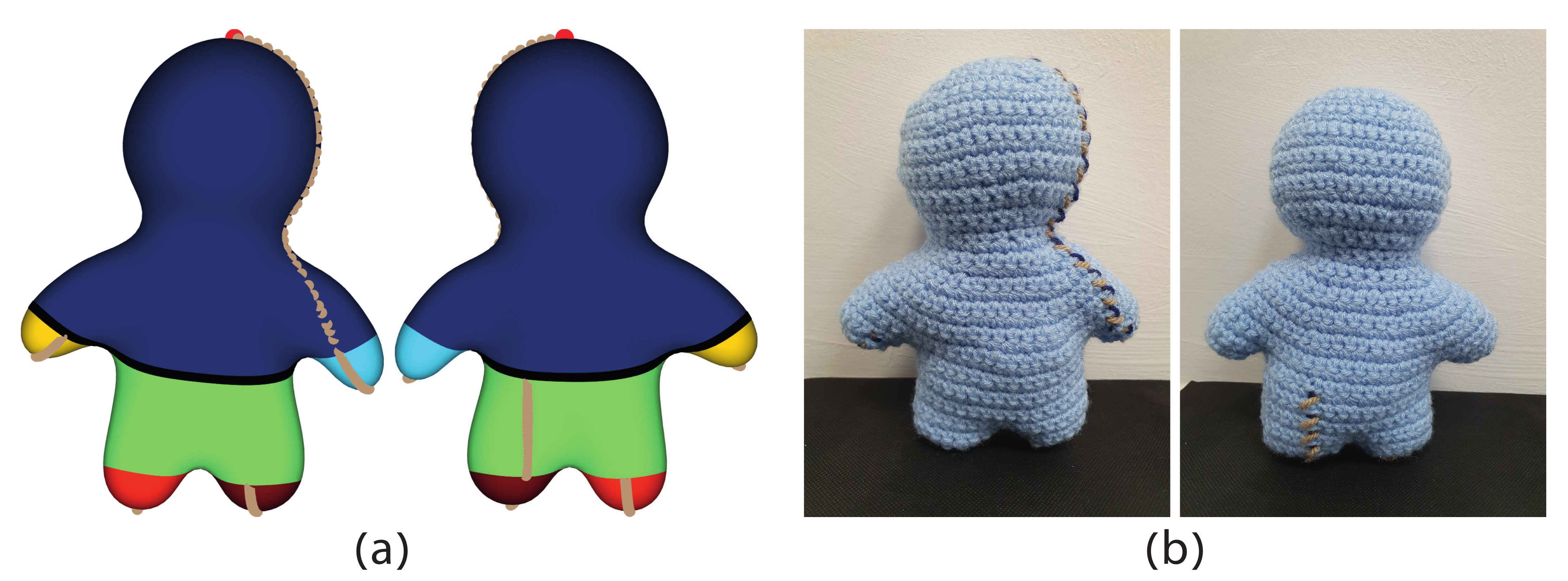}
		\caption{(a) The cut location (brown) on a segmented model. (b) The crocheted model with the cut marked using a piece of yarn.}
		\label{fig:man}
	\end{figure*}
	
	\subsection{Branching}
	\label{sec:multiple_segments}
	If for a given seed $\cs$, the geodesic function $f(x)=d(\cs,x)$ has multiply-connected isolines, the graph $\cm$ cannot cover the model. In these cases, the model is automatically decomposed into multiple segments, each of which can be crocheted using the approach described in Sections~\ref{sec:mesh2graph} and~\ref{sec:graph2pattern}. The segments are attached by a method called "join-as-you-go" \cite{JoiningAmigurumiLimbs}, meaning each segment is crocheted onto the last row of the previous segment, and therefore no additional sewing is required. Furthermore, the segment boundaries are not visible in the crocheted object. The more common method, in contrast, involves sewing together closed segments, which requires accuracy to achieve geometric details such as symmetry. In this section we describe the modifications required to accommodate such objects.
	
	\subsubsection{Mesh to Graph}
	Given a seed $\cs$ let $f(x)=d(\cs,x)$. Let $\sad=(\sigma_1,..,\sigma_m)$ be the saddle points of $f$, sorted by $f_i = f(\sigma_i)$. Namely $f_1 \leq f_2 \leq ... \leq f_m$. For each $\sigma_i$ in order, we compute the isoline of $f_i$, denoted by $\gamma_i$, and slice a new segment for each connected component of $\gamma_i$. Meaning, the segments are obtained by slicing along the isolines of the saddle points, ordered by increasing geodesic distance to the seed. Figure~\ref{fig:homer}(a) shows the isolines of $f$ for the seed point $\cs$ marked in red, as well as the saddles (cyan), and maxima (blue). 
	
	In addition to the segmentation, we generate a directed graph $G_\sigma$, whose vertices are the segments, and where an edge $(s,t)$ exists if the segments $\mM_s$ and $\mM_t$ share a boundary, and $f$ values on $\mM_s$ are smaller than $f$ values on $\mM_t$. The crocheting order of the segments is determined by a topological sort of $G_\sigma$.
	Figure~\ref{fig:homer}(b) shows the resulting segments. Very thin segments might not be sampled (marked in black in Figure~\ref{fig:homer}(b) and other figures), and are skipped and not crocheted.

	
	\paragraph{Geometry}
	Any resulting segment $\mM_l$ is either a half sphere or a cylinder, and thus can be covered by a crochet graph $\cm_l$. While $f$ is computed for the whole model before segmentation, $g$ is computed for each segment separately. 
	The cut for $g$ is made from the maxima of $f$ in the segment to the closest point on the segment's boundary. If $f$ attains its maximum on one of the boundaries of the segment (there are at most two boundaries), then cut is computed to the closest point on the other boundary.
	
	Figure~\ref{fig:man} shows an example of the location of the cut, where we show the front and the back of the model.
	We show the location of the cut in brown on the segmented model (a), as well as the location of the cut in the crocheted model (b). To mark the location of the cut during crocheting a piece of yarn was used to mark the beginning/end of the row.

	\paragraph{Connectivity}
	For every two segments $\mM_s,\mM_t$ which share an edge $(s,t)$ in $G_\sigma$, we add an additional condition that the last row of $\mM_s$ is coupled to the first row of $\mM_t$. 
	Figure~\ref{fig:homer}(c) shows the crochet graph for all the segments of the Homer model. Finally, (e) shows the crocheted model, where each segment was crocheted with a different color for better visualization.
	
	\subsubsection{Graph to instructions}
	
	The same simulation of the crochet operations is applied to the first row of a new segment,
	but the sequence of stitches that is used as its previous row is no longer a full row.
	Instead, the last rows of all attached segments are arranged and filtered to include only
	vertices that have a connecting edge to the new segment's row,
	constituting a ``joint'' previous row.
	The transducer then takes stock of when its consumption of the previous row skips stitches, splits segments, skips segments, or spans multiple parent segments, and includes this information in the row instructions.

	

	
	

	
	
	\section{Limitations}
	Our method only handles closed surfaces, since we aim for Amigurumi models, which are stuffed. 
	The segmentation approach may generate thin segments, which are harder to crochet. While we filter out very thin segments, we believe that small modifications to the singularity locations can yield a better segmentation without considerably affecting the shape.
	Our approach does not take the symmetry of the model into account, and thus discretization errors for low resolution patterns may lead to non-symmetric crocheted models.
	Our current setup also has limited design freedom. While using a single seed is simple, it does not give the user control of the knitting direction throughout the shape. 
	
	
	
	

	\section{Results}
	\subsection{Implementation Details}
	We implemented our algorithm in Matlab and C++. We use Heat Geodesics~\cite{crane2017heat} for computing geodesic distances, where we took the recommended time parameter $t$, namely the average edge length squared. In some cases, the mesh has too many geometric details, leading to neighboring saddles/extrema of the distance function. In this case, we take advantage of the tunability of heat geodesics, and repeated multiply $t$ by an increasing power of $2$ until there are no more neighboring saddles/extrema.
	For computing geodesic paths we use~\cite{sharp2020you}. Table~\ref{tab:models} provides the statistics for all the models. Algorithm running times were measured on a desktop machine with an Intel Core i7.
	
	\begin{table}
		\centering
		\caption{Statistics for the crocheted models. }
			\begin{tabular}{  l |  c| c |c |c } 
				
				Model & Rows & Segments & Stitches & Time (min) \\
				\hline
				
				Teddy, Fig~\ref{fig:teaser}, Fig~\ref{fig:gal} & 60 & 6 & \ME{3670} & 2.5 \\ 
				Teddy, Fig~\ref{fig:gal} & \ME{30} & \ME{6} & \ME{880} & \ME{2.1} \\
				Homer, Fig~\ref{fig:homer} & 50 & 6 & \ME{1605} & 2.0 \\ 
				Mushroom, Fig~\ref{fig:FBLO} & 30 & 1 & \ME{365} & 0.2 \\
				Man, Fig~\ref{fig:man} & 45 & 6 & \ME{1643} & 1.5 \\ 
				C, Fig~\ref{fig:gal} & 40 & 1 & \ME{1118} & 0.3 \\ 
				Fish, Fig~\ref{fig:gal} & 40 & 1 & \ME{1020} & 0.3 \\ 
				Bob, Fig~\ref{fig:gal} & 30 & 4 & \ME{854} & 1.0\\ 
				Bunny, Fig~\ref{fig:gal} & 45 & 5 & \ME{2470} & 0.8\\
				Moomoo, Fig~\ref{fig:gal} & 60 & 8 & \ME{2491} & 6.9 \\ 
				Pretzel, Fig~\ref{fig:gal} & 30 & 7 & \ME{586} & 1.5 \\ 
			\end{tabular}
			\label{tab:models}
		\end{table}

		\subsection{Gallery}
		Figures~\ref{fig:teaser},~\ref{fig:homer},~\ref{fig:leg_a} demonstrate our results. 
		Figure~\ref{fig:gal} shows additional models, where we show the (a) segmented shape and seed, (b) the crochet graph $\cm$ and its embedding $\cmX$ on the input mesh, (c) the embedding $\cmSU$ generated from the edge lengths, and (d) the final crocheted object.
		The sixth row of Figure~\ref{fig:gal} shows an example of a pretzel model, which cannot be scheduled for machine-knitting as discussed at AutoKnit~\cite{narayanan2018automatic} (see Figure 25 there). Our method, on the other hand, generates crochetable instructions.
		The last two rows of Figure~\ref{fig:gal} show the results for the same model, seed point, yarn type and hook but using different stitch width $\cw$. Note that while manually adapting the pattern to different sizes is a difficult task, our algorithm preforms it automatically.
		Note that the shapes are similar to the input, and the segment boundaries are not visible in the output crocheted models. Therefore, we can achieve varied geometries without visibly segmenting the shape, leading to more visually pleasing results than the approach by Igarashi et al.~\shortcite{igarashi2008knitting} (see Figure 11 there). Furthermore, our results have a closer resemblance to the input, compared to the method by Guo et al.~\shortcite{guo2020representing} (see Figure 12 there).

		
		
		
		
		\subsection{Creases}
		Since stuffed items tend to be smooth, there exist crochet shaping techniques that allow the generation of creases. Specifically, instead of inserting the hook under both loops of the stitch, the hook is inserted only in the front loop (denoted \emph{Front Loop Only} \pat{FLO}), or only in the back loop (denoted \emph{Back Loop Only} \pat{BLO}). The \pat{BLO} (resp. \pat{FLO}) stitch allows creases which are positively (resp. negatively) curved with respect to the columns direction (i.e. orthogonal to the knitting direction). 
		
\begin{figure}[t]
			\centering
			\includegraphics[width=\linewidth]{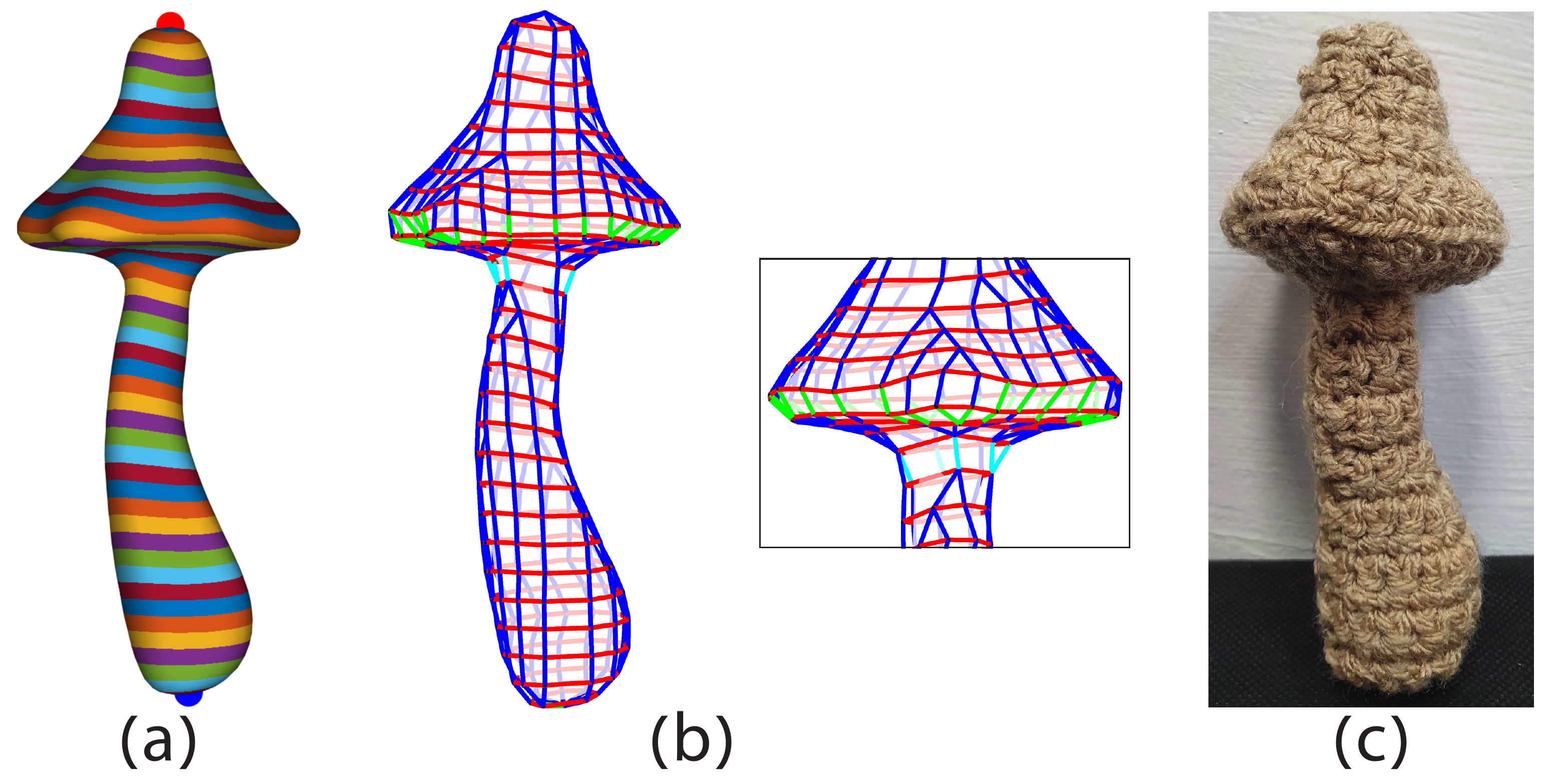}
			\caption{(a) The isolines of $f$ for the marked red seed point, and the maxima of $f$ (marked blue). (b) The  crochet graph with \pat{BLO} edges (green) and \pat{FLO} edges (cyan) (c) The final knitted model.}
			\label{fig:FBLO}
		\end{figure}

\begin{figure*}[b]
			\centering
			\includegraphics[width=0.7\linewidth]{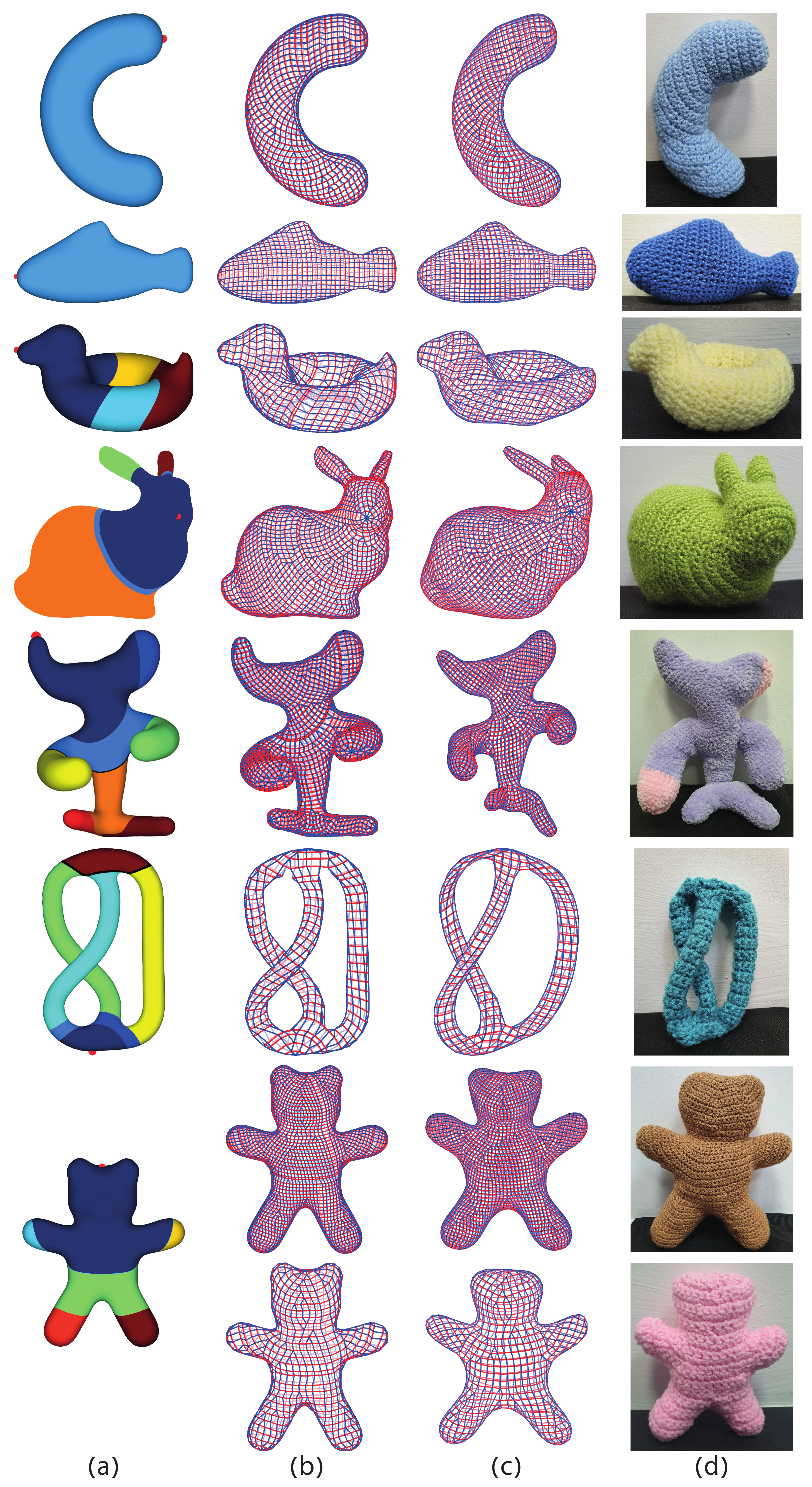}
			\caption{A gallery of our results. (a) The segmentation. (b) The crochet graph $\cm,\cmX$ (c) The embedding $\cmSU$ of the crochet graph. (d) The crocheted model. Note that the locations of the segments is not visible in the final output.}
			\label{fig:gal}
		\end{figure*}

		We define \emph{crease vertices} as vertices that have large maximum absolute curvature, and their maximum absolute curvature direction is orthogonal to the knitting direction. For any two consecutive crease vertices on the same row, we mark all the stitches based on these vertices as \pat{BLO} or \pat{FLO}, depending on the sign of the curvature.
		We allow the user to choose whether to enable this option or not. 
		Figure~\ref{fig:FBLO} shows an example of a model where this shaping technique was used. The \pat{BLO} (resp. \pat{FLO}) edges are marked in green (resp. cyan) in (b). Note the corresponding sharp crease in the knitted object (c). 
		We note that creases have also been used in knitting. For example, \cite{wu2019knittable} allow for creases, but they use a different technique.
		



		\section{Conclusions and Future Work}
		We presented a novel automatic approach for generating crochet knitting instructions for Amigurumi from an input triangle mesh. Given a single seed point and a stitch size, we generate human-readable instructions that use only simple crochet stitches (\pat{sc}, \pat{inc}, \pat{dec}). Our method is applicable to a variety of geometries, and leads to crocheted models which are visually similar to the input shape. In the future we plan to incorporate our method within an interactive framework that allows the user to move the seed point, change the yarn and the gauge and see the expected shape of the output. Furthermore, we plan to add colors and texture, as well as support for additional types of stitches.
		
		With the wide popularity of Amigurumi, and crochet in general, in recent years, we believe that tools that allow novice users, as well as pattern designers, to generate crochet instructions from 3D models would be quickly adopted and built upon by the crocheters and makers communities. Our approach provides an important stepping stone in this direction, and we expect that it will sow the seeds for further research in comptuational crochet.

		\begin{acks}
			M. Edelstein acknowledges  funding  from  the  Jacobs Qualcomm Excellence Scholarship and the Zeff, Fine and Daniel Scholarship. M. Ben Chen acknowledges the support  of  the Israel Science Foundation (grant No. 1073/21), and the European Research Council (ERC starting grant no. 714776 OPREP). We also thank SHREC’07 and Josh Holinaty for providing models.
		\end{acks}
		
		
		\bibliographystyle{ACM-Reference-Format}
		\bibliography{AmiGo}
		
		
		
	\end{document}